\documentclass{rspublic}
\usepackage{graphicx}

\usepackage{natbib}

\def\be{\begin{equation}}
\def\ee{\end{equation}}

\def\etal{{\it et al. }}

\begin{document}

\title[Twist stretch elasticity]{Strain dependent twist-stretch elasticity in chiral filaments}

\author[M.~Upmanyu \etal]{M.~Upmanyu$\mbox{}^{1,2,3}$,  H.-L.~Wang$\mbox{}^{1,2}$, H.-Y.~Liang$\mbox{}^{1,2,\ast}$ and R.~Mahajan$\mbox{}^4$}

\affiliation{$\mbox{}^1$ Group for Simulation and Theory of Atomic-Scale Material Phenomena ({\it st}AMP), Division of Engineering, Colorado School of Mines, Golden, CO 80401 \quad {\rm (mupmanyu@mines.edu)}\\
$\mbox{}^2$ Materials Science Program, Colorado School of Mines, Golden, CO 80401\\
$\mbox{}^3$ Bioengineering and Life Sciences Program, Colorado School of Mines, Golden, CO 80401\\
$\mbox{}^4$ Kennedy Krieger Institute, Baltimore, MD 21211\\
$\mbox{}^\ast$ Present address: Division of Applied Science and Engineering, Harvard University, Cambridge, MA 02138
}

\label{firstpage}

\maketitle

\begin{abstract}{chirality, supramolecular assemblies, mechanical coupling, Elastica, carbon nanotubes, DNA mechanics} 
Coupling between axial and torsional degrees of freedom often modifies the conformation and expression of natural and synthetic filamentous aggregates. Recent studies on chiral single-walled carbon nanotubes and B-DNA reveal a reversal in the sign of the twist-stretch coupling at large strains. The similarity in the response in these two distinct supramolecular assemblies and at high strains suggests a  fundamental, chirality dependent non-linear elastic behaviour.  Here, we seek the link between the microscopic origin of the non-linearities and the effective twist-stretch coupling using energy-based theoretical frameworks and model simulations. Our analysis reveals a sensitive interplay between the deformation energetics and the sign of the coupling, highlighting robust design principles that determine both the sign and extent of these couplings. These design principles have been already exploited by Nature to dynamically engineer such couplings, and have broad implications in mechanically coupled actuation, propulsion and transport in biology and technology. 
\end{abstract}

%\cite{asit:Castle:1942,asit:Preston:1950,asit:Green:1954}

\section{Introduction}

Structural chirality (i.e. handedness) often results in large mechanical couplings in filamentous supramolecular assemblies. As a classical example, twist-stretch elasticity in DNA is vital during  chromatin organization, transcription regulation and binding specificity during DNA-complex formation~(Hagerman 1988; Tsai \etal 1978; Hogan \& Austin 1987; Kornberg \& Lorch 1995; Mirkin 2000). The morphology of the helical, rotating flagella that provide thrust to several classes of bacteria during their chemotaxis is in part dependent on the sign and extent of such mechanical couplings~(Kitao \etal 2006; Coombs \etal 2002). They are also expected to influence the mechanics of twisted biopolymer aggregates such as coiled-coils~(Offer \etal 2002), amyloid fibrils~(Jimenez \etal 2002), and network forming fibers such as actin, fibrin~(Weisel \etal 1987; Medved \etal 1990) and collagen (Bozec \etal 2007). Similar chiral supermolecular filaments have also been synthesized, in particular at the nanoscale. In applications which rely on physico-mechanical couplings, engineering the coupled response in structurally robust helical nanowires~(Kondo \& Takayanagi 2000), nanobelt coils~(Pan \etal 2001), chiral nanotubes, and their assemblies offers an elegant route for fabrication of nanoscale motors, oscillators and switches by
%, twist-stretch elasticity 
eliminating the need for externally actuated deformation~(Craighead 2000; Baughman \etal 1999; Baughman \etal 2002; Fenimore \etal 2003; Papadakis \etal 2004; Liang \& Upmanyu 2005; Liang \& Upmanyu 2006; Cohen-Karni \etal 2006).

Twist-stretch coupling in chiral, elastically stiff filaments is usually described by
%When the persistence length is much larger than the scale associated with the mechanical coupling, the filament is elastically stiff (elasto-filament). 
linearized Elastica~(Love 1944). Well below the relevant persistence lengths, the effect of thermal fluctuations is negligible and the free energy per unit length of a straight filament subject to an axial force $F$ can be expressed in terms of the strain components,
\begin{equation}
\label{eq:EnergyFunc}
\mathcal{U} = \left[ \delta U_b(R) + \frac{1}{2} k_\epsilon \epsilon^2 + \frac{1}{2} k_\phi \phi^2 + k_{11}\;\phi\epsilon + \ldots \right] - F\epsilon.
\end{equation} 
%$\mathcal{F} \propto \left[  \left( k_\epsilon \; \epsilon^2 + k_\phi \; \phi^2 + k_{\epsilon\phi} \; \phi \epsilon + \ldots \right) - F \epsilon \right]$
%consider the free energy function   of a filament subject to an axial force $F$, 
 %where  and
%The first term accounts for the change in rod radius (with $k_R$ the associated rigidity), 
%\frac{1}{2}k_R\, \delta\left(\frac{1}{R^2}\right) +
The first term is the contribution, if any, due to the change in radius of the filament. For a hollow tube of radius $R$, this contribution can be expressed as a change in the bending energy of the rolled-up sheet due to its mean curvature $\kappa=1/R$, i.e. $U_b = \frac{1}{2} k_b/R^2$ with $k_b$ the bending rigidity of the rolled up sheet. The constants $k_\epsilon$ and $k_\phi$ are the extensional and torsional rigidities of the rod, while $k_{11}$ couples the twist per unit length $\phi$ to axial strain $\epsilon$.  The magnitude of $k_{11}$ is a measure of the interplay between structural chirality and the microscopic deformation mode. 

Minimizing the free energy at a fixed applied strain ($\epsilon$ or $\phi$) yields the equilibrium conjugate strain, and therefore the effective coupling. As an example, for negligible change in the filament radius, the coupling reduces to
\begin{equation}
\label{eq:eqStrains}
\phi_{eq} = - \frac{k_{11}}{k_\phi} \epsilon, \quad \text{and}\quad \epsilon_{eq} =  - \frac{ k_{11}\phi - F}{k_\epsilon},
\end{equation} 
where $\phi_{eq}$ and $\epsilon_{eq}$ are the equilibrium strains induced by prescribed conjugate strains $\epsilon$ and $\phi$, respectively. Similar energy-based models have been used to describe the coupled mechanical response of B-DNA~(Marko 1997; Kamian \etal 1997; Bustamante \etal 2000; Lionnet  \etal 2006), twist-storing semiflexible biopolymers~(Moroz \& Nelson 1998), chromatin fibers~(Ben-Haim \etal 2001), and bacteria flagella~(Coombs \etal 2002).

%Minimizing the free energy at a fixed applied strain ($\epsilon$ or $\phi$) yields the equilibrium conjugate strain, and therefore the effective coupling. Such energy-based models have been used to described the coupled response in B-DNA~[\cite{biofil:Marko:1997,biofil:KamianNelson:1997,biofil:Bustamante:2000,biofil:BensimonCroquette:2006}], twist-storing semiflexible biopolymers~[\cite{biofil:MorozNelson:1998}], chromatin fibers~[\cite{biofil:HaimLesne:2001,biofil:BancaudPrunell:2006}], and bacteria flagella~[\cite{biofil:CoombsGoldstein:2002}].

This article is motivated by the recent studies on axially strained chiral single-walled carbon nanotubes (SWCNT) and B-DNA. Surprisingly, the effective coupling $d\phi/d\epsilon$ exhibits a sign reversal in both filaments, past a critical strain $\epsilon_c$ - under compression in chiral SWCNTs~(Liang \& Upmanyu 2006) and in stretched B-DNA~(Gore \etal 2006). 
%There is however a difference in the sign and evolution of the coupling. Compared to a helical spring, the SWCNT response is intuitive at small strains, while opposite is the true for B-DNA. 
Evidently, the strain dependent coupled response cannot be described by linearized Elastica (equations~\ref{eq:EnergyFunc} \& \ref{eq:eqStrains}). In an attempt to understand and quantify this behaviour, we isolate the microscopic response in each case, which then serves as the physical basis for developing energetic frameworks that capture the non-linear response.
 
%The sign reversal of tiwst-stretch coupling in reveal that  develops a strain dependence. In both cases, the coupling 
\section{SWCNT Response}
The atomic-scale simulations on the SWCNTs were performed using the second generation reactive bond order (REBO) potential, which reproduces several mechanical properties of fullerenes, graphenes and diamond to a high degree of accuracy~(Brenner \etal 2002). The results discussed here are limited to the (8,4) chiral nanotube, i.e. the SWCNT axis is set by the chiral vector $C_h=8 \textbf{a}_1 + 4 \textbf{a}_2$, where $\textbf{a}_1$ and $\textbf{a}_2$ are the graphite basis vectors. A Langevin bath was used to fix the temperature ($0.1^\circ$K) and the strain was applied in increments of 0.25\% by displacing the end atoms (shown masked in figure~1A). A conjugate gradient relaxation was performed between each successive strain step. During relaxation, the end atoms were allowed to displace along the tangential and radial directions. The reported twist (per unit length) is the relative rotation between the SWCNT ends averaged over the net rotation of the end atoms. For details, see Liang \& Upmanyu (2006).
\begin{figure}[htb]
\centering
\includegraphics[width=10.0cm]{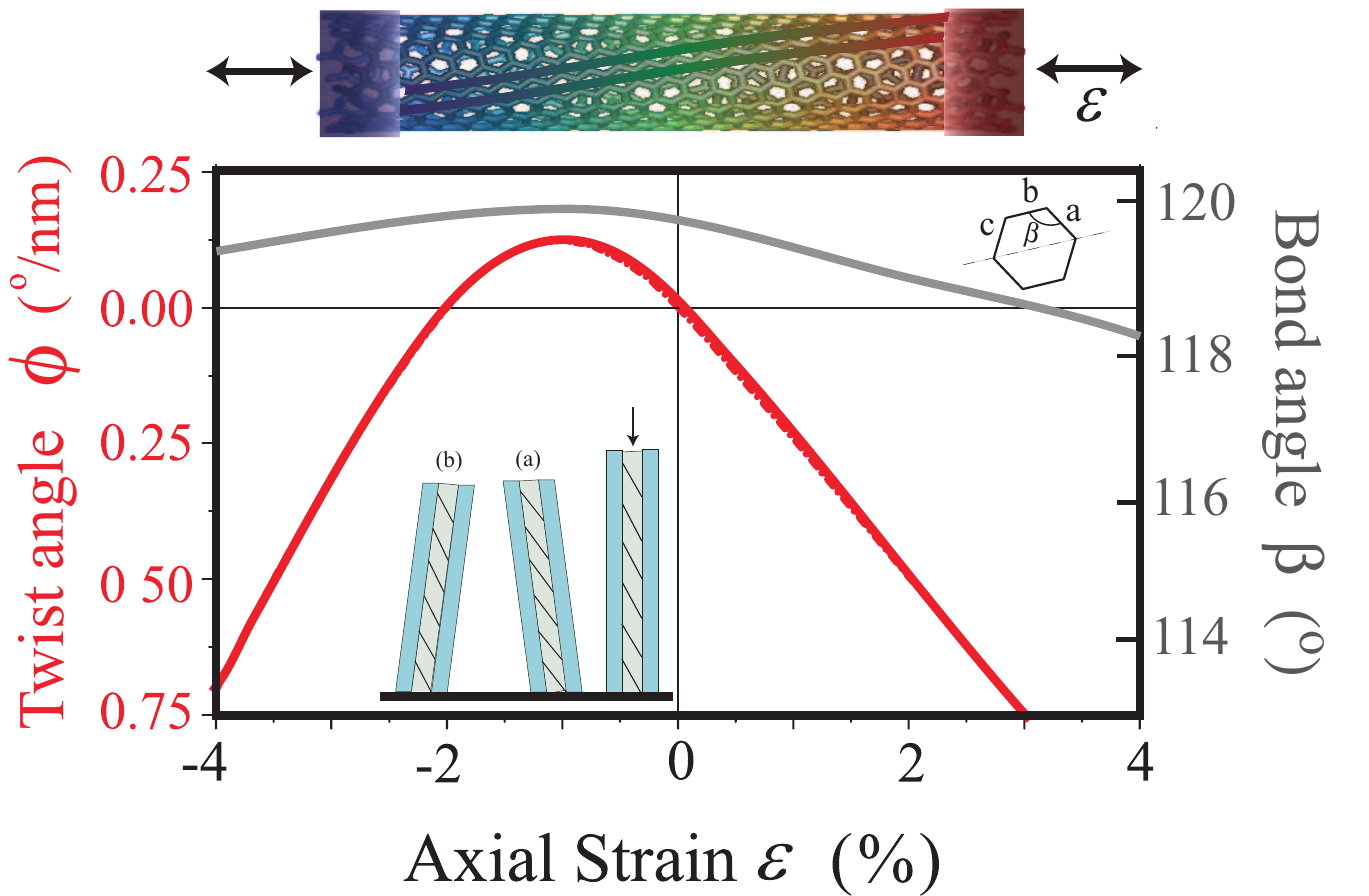}
\caption{{\bf Twist stretch elasticity in a chiral SWCNT}: (top) Axial distribution of relative twist (color scale) in an axially strained chiral (8,4) SWCNT~(Liang \& Upmanyu 2006). The helical arrangement of the graphene structural units is shown via bold lines. (bottom) The strain dependence of the coupling (red line) and one of the three unique bond angles $\beta$ (gray line), both of which exhibit a reversal at $\epsilon\approx2\%$. (bottom left inset) Schematic illustrating the role of inter-fiber shear instability in reversing the coupling between axial and shear strains. The fibers connecting the two filaments can deform via fiber extension and inter-fiber shear (slip). In (a), the compression of the filaments proceeds via a combination of both, resulting in reorientation of the fibers and shear to the left. The shear is reversed in the case of enhanced inter-fiber with negligible fiber rerientation, as illustrated in (b).
\label{fig:SWCNTResults}}
\end{figure} 

The sign of the SWCNT coupling constant $k_{11}$ is positive at small strains and consistent with the SWCNT handedness (see figure~\ref{fig:SWCNTResults}), the chiral angle between the SWCNT axis and the basis vectors of the underlying graphitic lattice. The ``twist-hardening" under compression forces the nanotube to untwist and then twist in the opposite direction. The change in its radius is negligible during this reversal, consistent with its relatively small Poisson's ratio (Yakobson \etal 1996; Jin \& Yuan 2003), $\nu\sim0.2$. The extensional and torsional rigidities are weak functions of the SWCNT chirality~(Vaccarini \etal 2000, Jin \& Yuan 2003), ruling out the direct effect of changes in chirality due to the induced twist. 

A potential source for the non-linear response is the chirality dependence of the coupling constant, $k_{11} = k_{11}^0 \sin6(\theta_0\pm\phi)$, where $\theta_0$ is the chiral angle of the relaxed SWCNT and $k_{11}^0$ is a constant~(Liang \& Upmanyu 2006). This specific form satisfies the symmetry constraints, in particular the six fold symmetry of the underlying graphitic lattice, and leads to terms in higher powers of $\phi$ and a strain dependent coupling, 
\begin{equation}
k_{11} = k_{11}^0 \sin6\theta_0 \pm 6\phi \cos6\theta_0 - 18\phi^2\sin6\theta_0 + \ldots. 
\end{equation}
However, the effect of these higher order terms is small for the high chirality SWCNTs for which the coupling has been observed ($\theta_0\sim\pi/6$) as $\cos6\theta_0\sim0$ and the ratio $\epsilon/\phi\sim10$ over the sign reversal range, effectively ruling out chirality dependence of the coupling constant as the basis for the observed strain dependent coupled response.

Microscopic analyses of the strain dependence of the bond lengths and bond angles within the simulations reveal that the reversal coincides with that in the trend associated with one of the three bond angles between the carbon atoms on the SWCNT surface (angle $\beta$, see figure~\ref{fig:SWCNTResults}). This change in the mechanism of strain accommodation effectively softens the nanotube surface with respect to shear, which modifies the otherwise affine response of the graphene network. The helically coiled graphene elements slide past each other and the enhanced surface shear counteracts the tendency to increase the SWCNT handedness. An analogous, two-dimensional response is illustrated schematically in figure~\ref{fig:SWCNTResults} (inset, bottom left) for a sheet with bare elastic anisotropy which now couples the compression to its shear. The deformation of the anisotropic layer responsible for the coupled response, in particular the ratio of layer compression to intra-layer slip, controls the direction of the shear.

The basis for this effect is an additional elastic anisotropy term in equation~\ref{eq:EnergyFunc} that we have ignored so far because it is weakly chirality dependent~(Gartstein \etal 2003),
\begin{equation}
\delta \mathcal{U}(\theta, \phi, \epsilon) \propto (\epsilon^2 - \phi^2) \cos6(\theta_0+\phi).
\end{equation}
Cleary this term becomes important at large axial strains, with the leading contribution of the form $k_{12} \phi \epsilon^2$, with $k_{12}$ a constant. Incorporating this term and maximizing the resultant energy functional, we recover a strain dependent  equilibrium twist and the critical axial strain associated with the twist reversal, 
\begin{equation}
\phi_{eq} = - \frac{\epsilon k_{11} +\epsilon^2 k_{12}} {k_\phi},\quad \epsilon_c = -\frac{k_{11}} { 2 k_{12}}.
\end{equation}
For the (8,4) SWCNT, $k_\phi = 280$\,nN\,nm$^2$, $k_{11} = 100$\,nN\,nm and $\epsilon_c=-1.3\%$~(Vaccarini \etal 2000; Liang \& Upmanyu 2006), which yields the higher order coupling constant, $k_{12} = -3.9\,\mu$N\,nm.
\begin{figure}[htb]
\centering
\includegraphics[width=10.0cm]{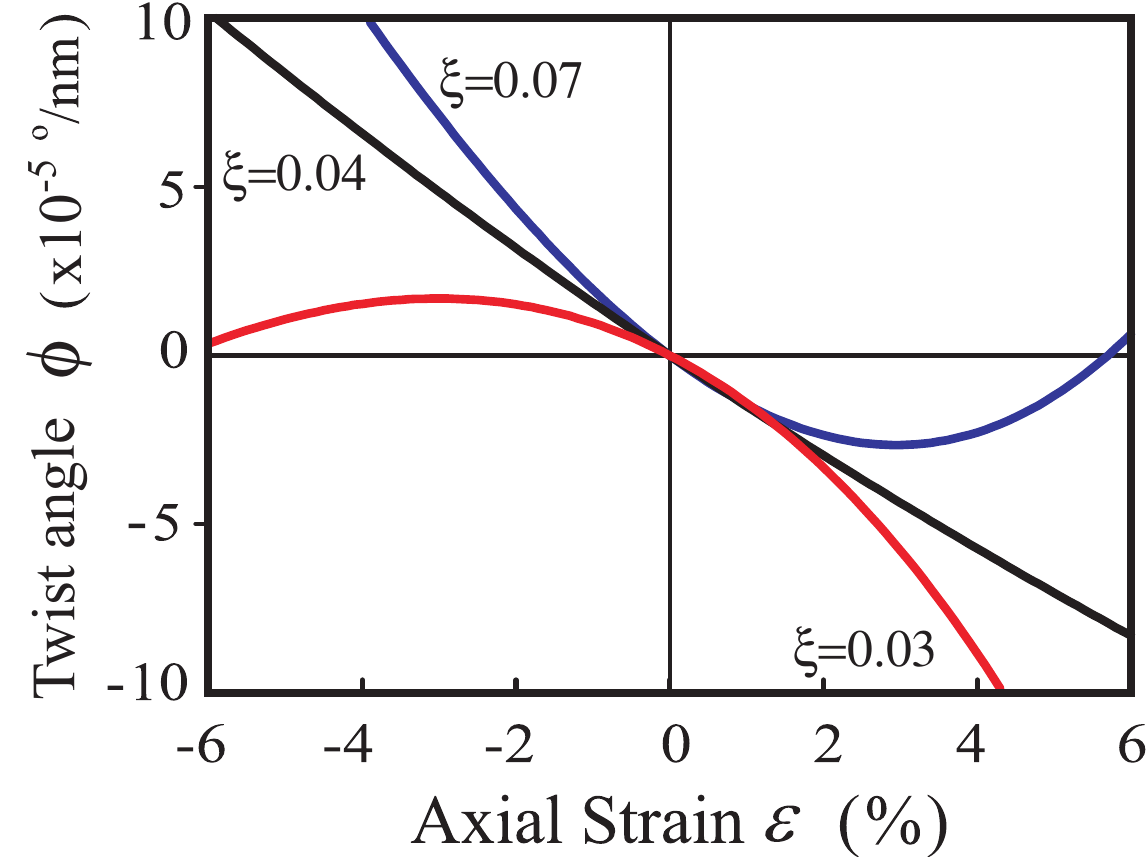}
\caption{{\bf Twist stretch elasticity in a model chiral nanotube}: Finite element simulations of twist-stretch elasticity in a model chiral filament formed by rolling a hexagonal lattice of I-shaped beam elements with varying beam geometries, as parameterized by $\xi=I/(AL^2)$.
\label{fig:FEMResults}}
\end{figure}

The strain dependent coupling points to an elegant design principle whereby the imposed strain can be used to engineer the coupling by effecting transitions in the microscopic deformation. Can we then fashion the structure so as to evoke the desired coupled response?  As an answer, we study the coupled response in an identically structured 
%system with highly simplified interactions. In this 
model (8,4) chiral nanotube with simplified interactions. The carbon-carbon bonds are replaced by I-shaped beam elements of equal length ($L=0.142$\,nm). The area of cross section is fixed at $A=0.005$\,nm$^2$. The Young's modulus and Poisson's ratio of the beam element material are fixed at $E=1$\,GPa  and $\nu=0.2$. These elements can either stretch or bend in order to absorb the imposed axial strain; the twisting of the beams is negligible and ignored.  The energy stored in bending and stretching scales quadratically with the deformation variables, $\epsilon$ and the beam curvature $\kappa$. The shape of the beams is important as it allows us to independently control the energetic competition between beam bending and stretching, the dimensionless ratio $\xi=I / (AL^2)$, where $I$ is the moment of inertia of each beam.
\begin{figure} [htb]
\centering
\includegraphics[width=10.0cm]{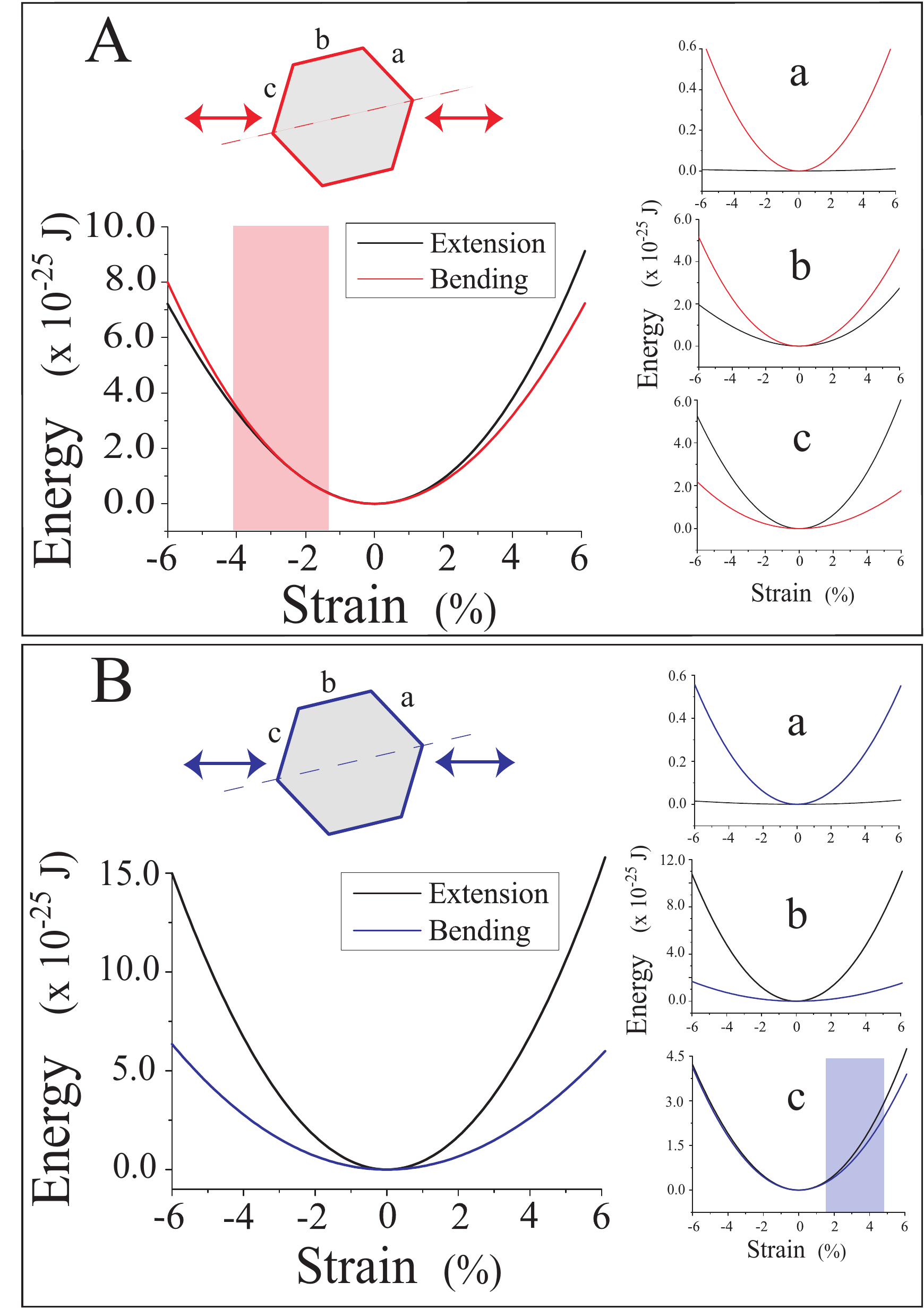}
\caption{{\bf Beam energetics in the finite element simulations}: Strain dependence of the energy stored in beam bending and stretching for $\xi=0.03$ (A) and $\xi=0.07$ (B) model nanotubes. The individual beams $a$, $b$ and $c$ for which the strain are calculated (right) are shown in the inset. The double arrows indicate the direction of axial strain. For clarity, the strain interval over which the reversal in the coupled response occurs is shaded.}
\label{fig:FEMAnalysis}
\end{figure}

The software package ANSYS is used to perform the finite element simulations of the strained filament. A non-linear solver is employed to ensure that the errors at high axial strains are minimal. The axial strain and measurement of the induced twist is performed as in the SWCNT case. In addition, strain dependence of the deformations of the individual beams are also monitored.  Simulations for varying values of $\xi$ indicate that the small strain coupling is consistent with the chirality, i.e. the twist changes sign with the axial strain (figure~\ref{fig:FEMResults}). Remarkably, the large strain response is sensitive to the value of $\xi$ and is quite rich. It varies from no reversal ($\xi=0.04$), to reversal under compression ($\xi=0.03$) and even tension ($\xi=0.07$). 

In order to gain a more rigorous understanding of the $\xi$ dependence of the coupled response, we have analysed the microscopic energetics of the individual beam elements as a function of the applied strain. Figures \ref{fig:FEMAnalysis}A ($\xi=0.03$) and \ref{fig:FEMAnalysis}B ($\xi=0.07$) show plots of energies associated with bending and stretching the beam elements for the entire filament  as well as for the three unique beam elements ($a$, $b$ and $c$) within a unit hexagonal cell on the filament surface. For the low $\xi$ filament, change in the coupling sign (shaded region) follows a gradual change in the energetic competition between bending and stretching under compression, with main contributors to this change being beams $b$ and $c$. A similar transition is observed under tension for the high $\xi$ filament. The balance in the energy stored in beam $c$ distinctly shifts towards stretching at about the same strain level at which we see the reversal in the coupling for the entire filament (shaded region). Finally, no such reversal in trends is observed for the intermediate $\xi$ ($=0.04$) filament (not shown), where the coupling more or less remains constant at all strains (figure~\ref{fig:FEMResults}).

The fact that this simplified model captures the essence of the SWCNT response indicates that the behaviour is both scale and model independent. The variations in the coupling are a direct reflection of the transitions in the microscopic energetic balance, quite like the observed surface shear instability in SWCNTs. 
%The low $\xi$ filament exhibits a transition under compression (beams $b$ and $c$, figure~S1), while a similar reversal is observed under tension in the high $\xi$ filament (beam $c$, figure~S2). 
Each transition disrupts the trends in the microscopic distortions, changing the sign of the coupling. While the small strain response is controlled by the chirality, the strain dependence of the coupling arises due to atomic-scale transitions in the deformation modes. The effect  is captured in an Elastica-based framework by a higher order coupling constant $k_{12}(\xi)$ which depends on the microscopic energetic balance and therefore modifies the energetic coupling between stretching and twisting the filament.

\section{B-DNA Response}
\begin{figure} [htb]
\centering
\includegraphics[width=10.0cm]{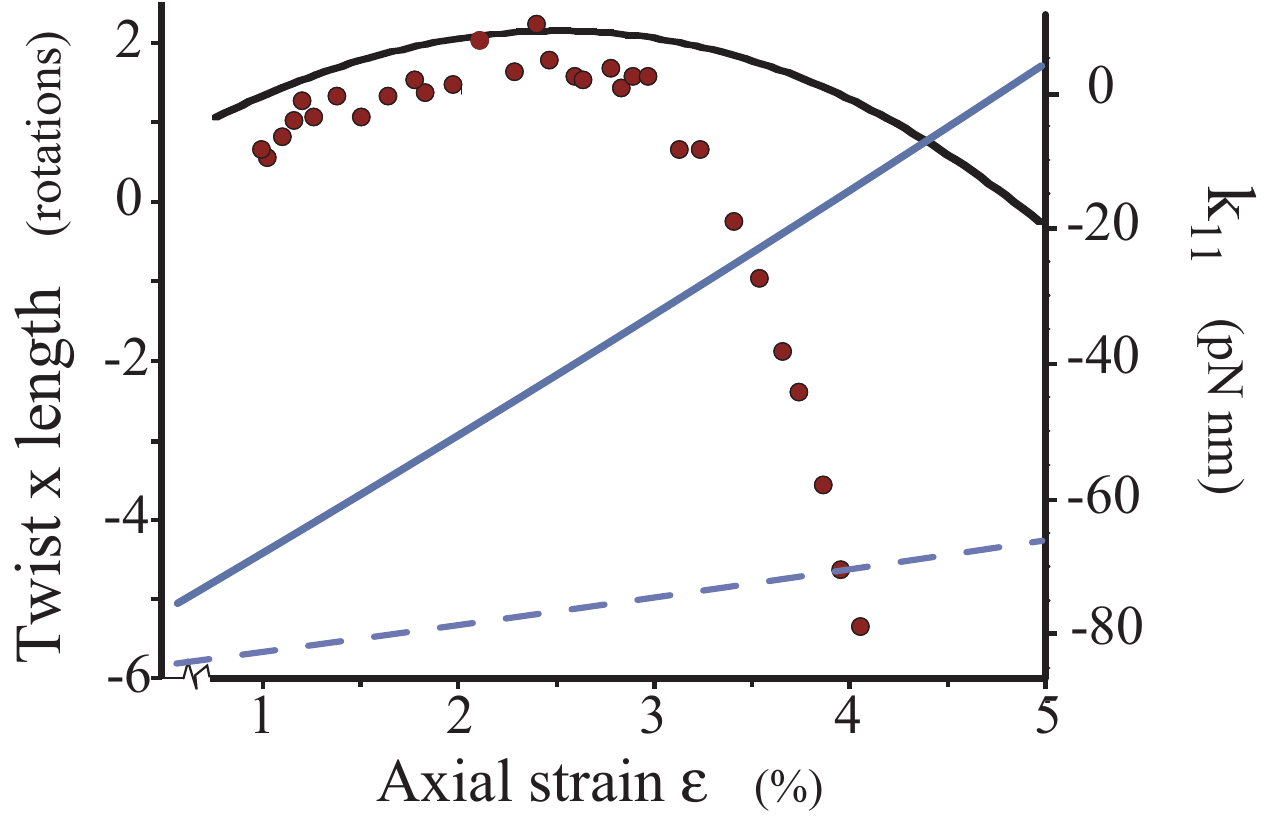}
\caption{{\bf Twist stretch elasticity in B-DNA} (left): The strain dependence of the twist-stretch coupling predicted by Eq.~2 (solid black line). The behaviour observed in bead assay experiments on a 8.4 kb long DNA molecule is also plotted for comparison (solid circles, Gore~\etal 2006). (right) Strain dependence coupling constant $k_{11}$, predicted by our analysis (solid line) and that by the toy model employed by Gore~\etal (2006).
\label{fig:DNAResults}
}
\end{figure}
In B-DNA, the sign reversal of the effective twist-stretch coupling observed in rotor bead experiments by Gore~\etal (2006) on sufficiently long individual molecules occurs past the entropic elastic regime.  The bend fluctuations are frozen and the molecule behaves like a chiral elastic rod~(Kamian \etal 1997; Bustamante \etal 2000). The critical strain associated with the reversal is well below that corresponding to the force at which the molecule undergoes a structural transition~(Smith \etal 2006; Bustamante \etal 2000; Lionnet~\etal 2006; Wereszczynski \& Andricioaei 2006). The microscopic basis for sign reversal appears to be different from that in the SWCNTs as the response may involve a substantial decrease in the DNA radius owing to its large Poisson's ratio ($\nu\sim0.5$), i.e. $R\approx R_0 (1-\nu\epsilon)$ with $R_0$ the relaxed effective radius~(Gore \etal 2006; Lionnet \etal 2006).

Then, when the extension is prescribed, the coupled response is modified due to radius dependence of torsional rigidity $k_\phi$, in addition to that of the coupling parameter $k_{11}$. Ignoring the direct effect of chirality change, the torsional rigidity scales as $R^4$. For short-range elastic interactions, scaling and symmetry considerations require that for harmonic distortions the coupling $ k_{11}$ constant is an odd function of $R$, i.e. $k_{11} (R) = \sum_{n\ge0} K_{2n+1} / R^{2n+1}$, where $K_{2n+1}$ are constants. In general, these constants will also be function of the helicity of the DNA which may change with the change in radius. We ignore this relatively smaller effect. We further assume that the double-stranded DNA behaves as a chiral rod composed of a material with bare surface anisotropy, i.e. the surface shear associated with the twist $\tau(\propto\phi_{eq} R)$ is non-zero for $R\rightarrow\infty$. As comparison, note that in the case of SWCNTs, the leading order term must be $1/R^3$ as the coupling is purely curvature derived, yet the graphite sheet is isotropic~(Liang \& Upmanyu 2006). This is a generalization of the toy model assumed by Gore \etal (2006), where the chirality-dependent small strain response was based on stiffer outer helical wire encompassing a compressible, isotropic inner core. 

Minimizing the energy $\partial\mathcal{U}/\partial \phi$ at constant axial strain $\epsilon$, we get
\begin{equation}
\phi_{eq} = -\frac{\epsilon}{k_{\phi}(R)} k_{11}(R) = - \frac{\epsilon}{k_\phi(R_0)} \left(\frac{R_0}{R}\right)^4 \left( \frac{K_1}{R} + \frac{K_3}{R^3} + \ldots \right),\nonumber
\end{equation}
where $R_0$ is the effective radius for the unstrained double strand ($\epsilon=0$). Plugging in the strain dependence of the radius $R=R_0(1-\nu \epsilon)$, 
%\[
%\phi_{eq} (\epsilon)= - \frac{\epsilon}{k_\phi(0)} \left[ \frac{K_1}{R_0} (1-\nu \epsilon)^{-5} + \frac{K_3}{R_0^3} (1-\nu \epsilon)^{-7} + \ldots \right] 
%\]
a series expansion limited to $\mathcal{O}[\epsilon]^3$ yields the induced twist,
\begin{equation}
\phi_{eq} (\epsilon) = - \frac{\epsilon}{k_\phi(0)} \left( k_{11}^\ast + k_{12}^\ast   \epsilon  + k_{13}^\ast \epsilon^2 + \mathcal{O}[\epsilon]^3 \right),
\end{equation}
with $k_{11}^\ast = k_{11}(0)$, and $k_{12}^\ast$ \& $k_{13}^\ast$ functions of $R_0$, $\nu$, $K_1$ and $K_3$.
%\begin{equation}s
%\phi_{eq} (\epsilon)= \frac{\epsilon}{k_\phi(0)} \left[\frac{K_1}{R_0}(1-\nu\epsilon)^{-5} + \frac{K_3}{R_0^3}(1-\nu\epsilon)^{-7}+\ldots \right].
%\end{equation}
Using the known mechanical properties of B-DNA ($R_0=0.924$\,\AA, $k_\phi(0)=460$\,pN\,nm, $k_{11}(0)=-90$\,pN\,nm) and the value of the critical strain associated with sign reversal, $\epsilon_c= 2.5\%$, we get $K_1=1761$\,pN\,nm$^2$ and $K_3=-1433$\,pN\,nm$^4$~(Bustamante \etal 2000; Gore \etal 2006). The predicted equilibrium twist as a function of the stretch is in excellent agreement with the experimental data, plotted in figure~\ref{fig:DNAResults}. However, the agreement rapidly breaks down past the critical strain, possibly due to increasing anharmonicity in the deformation~(Wiggins \etal 2006) and/or an increase in the chirality dependence of the coupling constant.

The analysis reveals that the magnitude of coupling constant (also plotted) decreases more rapid than that predicted by Gore~\etal (2006). In effect, stretched B-DNA always decreases its chiral character. While the molecular strain accommodation mechanism is yet to be established, it manifests as a decrease in the effective radius of the molecule and the molecule overwinds initially. Thereafter, the balance tilts in favour of the chirality dependent response such that the coupling changes its sign and then the molecule begins to unwind. 
%\[
%k_{11} = -90 + 1771.3 + 2247.9 \epsilon^2 + \mathcal{O}[\epsilon]^3,
%\]
%Here, $K_0$ is a measure of the bare anisotropy, i.e. the coupling for an ``untwisted" double strand. Since, the coupled effect is curvature derived, $K_0=0$.

\section{Discussion and Conclusions}
The SWCNT and B-DNA response both highlight the link between the microscopic strain accommodation behaviour and the macroscopic coupling. In the SWCNT case, we have clear evidence that the interplay between chirality and geometry is modified by the microscopic energetics, a general design principle for tailoring the extent and sign of the coupled response. This could be achieved by changing the chemistry  of the nanofilament, for example, and has direct relevance for tailoring the response of helical nanowires~(Kondo \& Takayanagi 2000; Tosatti \& Prestipino 2000), coiled nanobelts~(Pan \etal 2001),  and nanotube ropes~(Liang \& Upmanyu 2005). In each case, the extent of the response can modify the electromechanical couplings that can be easily exploited in NEMS and nanoelectronics. Our analysis on B-DNA shows that the same design principles are also important in semiflexible, soft chiral filaments. The change in radius of the stretched filament is likely to be controlled by the stiffness of the sugar phosphate backbone relative to that of the base-pair interactions across the strands. Similar energetic balance may also be responsible for the twist-stretch coupling recently extracted in chromatin strands~(Bancaud \etal 2006), although structural transitions may play a more important role there. Identification of such energetics is critical for understanding the coupled response in fibrillar assemblies such as coiled-coils and twisted protein aggregates. 
\begin{figure} [htb]
\centering
\includegraphics[width=10.0cm]{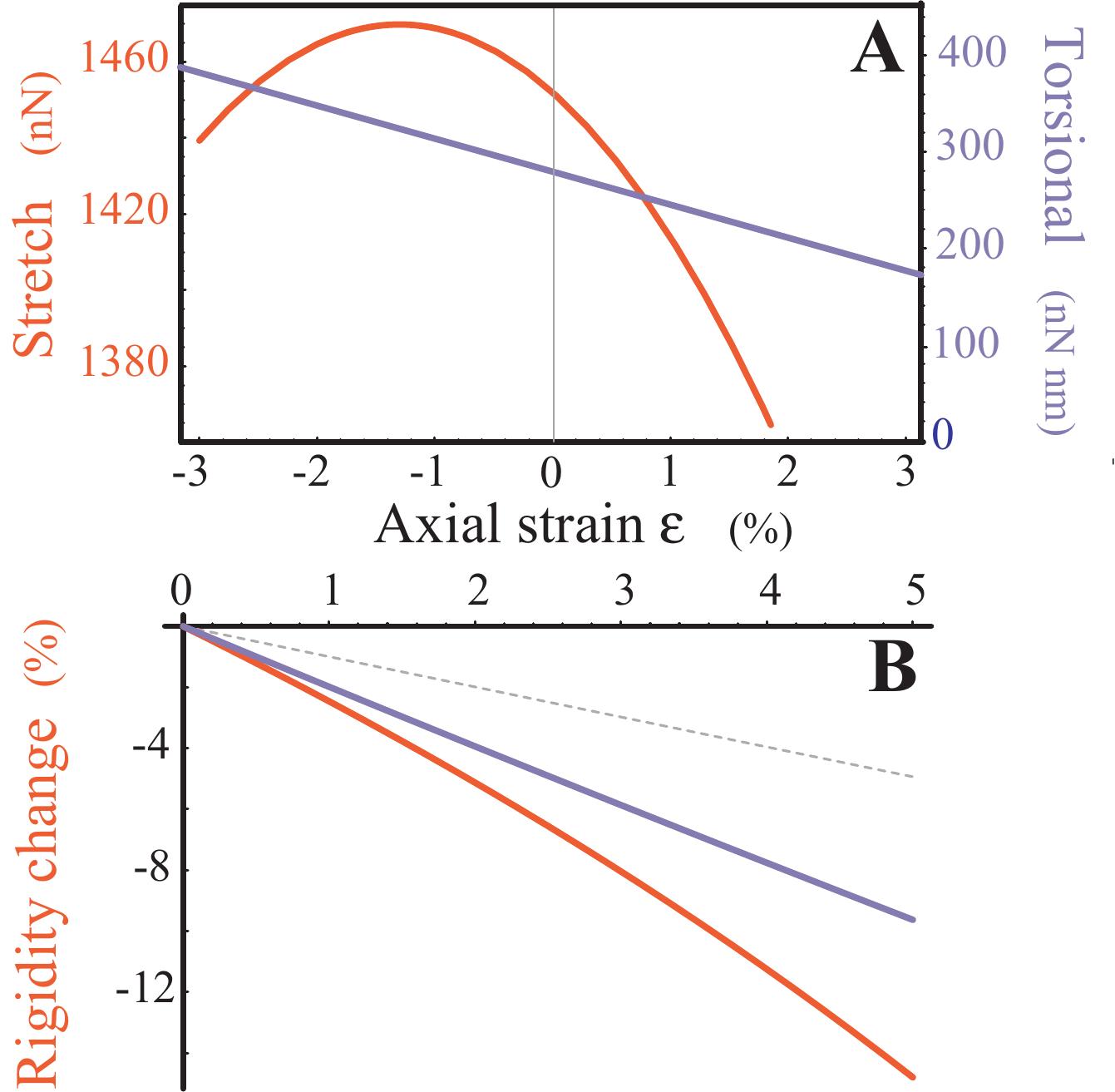}
\caption{{\bf The influence of twist-stretch coupling on effective rigidities}. Strain dependence of stretch (left) and torsional (right) rigidities for (A) the chiral SWCNT in figure~\ref{fig:SWCNTResults} and (B) B-DNA.
\label{fig:rigData}}
\end{figure}

An important consequence of the non-linearities is that they also modify the effective rigidity and electromagnetic character of these elastic filaments. The extensional rigidity of the (8,4) SWCNT develops an additional dependence on the induced twist, 
\begin{equation}
\frac{\partial^2\mathcal{U}}{\partial\epsilon^2} = k_\epsilon + 2 k_{12}\, \phi_{eq}.
\end{equation}
As a result, the nanotube softens rapidly as it is stretched and is the stiffest when compressed to the critical strain. A $2\%$ stretch induces a $7\%$ softening. The effective torsional rigidity $\partial^2\mathcal{U}/\partial\phi^2$ changes primarily due to the twist induced change in chirality. This effect is substantial - a 2\% axial strain leads to a 20\% change in the rigidity (see figure~\ref{fig:rigData}A). The reduction the stretch rigidity is also seen in B-DNA, where the radius dependence of the rigidities ($k_\epsilon\propto R^2$ and $k_\phi\propto R^4$) is additionally modified by that of the coupling constant $k_{11}$ (figure~\ref{fig:rigData}B), an effect that further amplifies the unusually large ratio of the twist to stretching/bending rigidity at high strains. Similar softening may also be responsible for the large  stretching strains that can be absorbed by twisted fibrin polymers~(Liu \etal 2006), a crucial requirement for the overall viscoelastic properties of blood clots.

The electromagnetic character of these nanofilaments must depend on the sign and extent of these couplings. It is well-known that the SWCNT band gap can be tuned by uniaxial and torsional deformation~(Minot \etal 2003). Close to the Fermi level, the change in band-gap is the highest for achiral nanotubes. The presence of this natural coupling in chiral SWCNTs, ignored in these analyses, should increase the efficiency of strain engineered band gaps (MU, in preparation). While there is considerable debate on the electronic properties of DNA~(Dekker \& Ratner 2001), it is clear that the the inherent coupling must be accounted for in understanding the interplay between their flexibility and electronic transport.

\begin{figure} [htb]
\centering
\includegraphics[width=10.0cm]{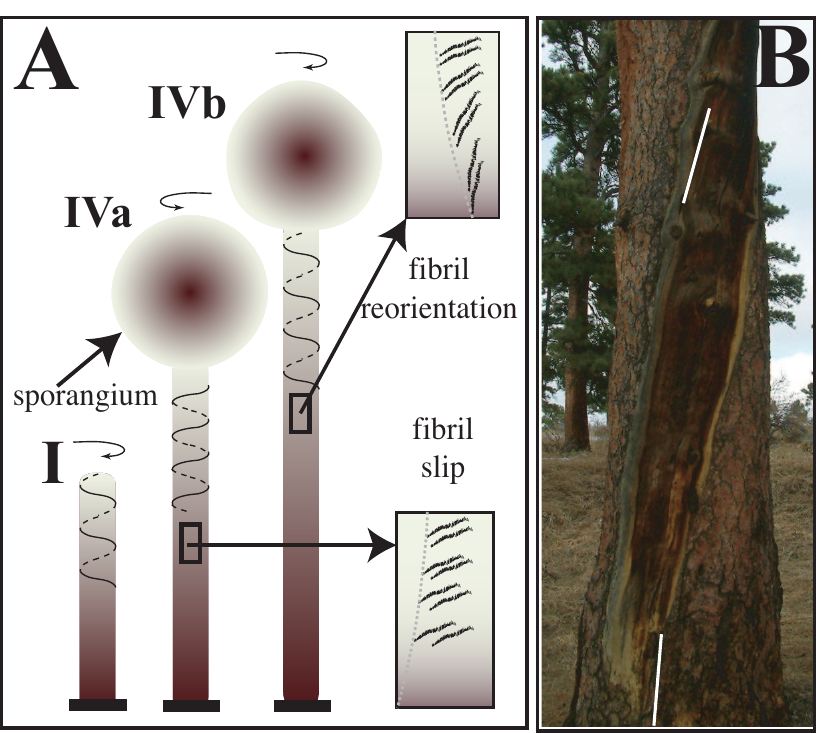}
\caption{{\bf Natural examples of engineered reversals in twist-stretch coupling}: (left) Schematic of spiral growth reversal in late stage (IVa-b) growth of {\it Phycomyces blakesleeanus}. (inset) Illustration of the rotation reversal due to the transition from fibril slip to fibril re-orientation (Ortega \& Gamow 1974).  (right) Spiral grain variations along the stem in a lodgetree pine ({\it Pinus contorta}). The white markers are visual guides for identifying the change in grain spiral. \label{fig:natEg}}
\end{figure}
There are several similar examples where Nature has dynamically engineered such mechanical couplings. Spiral growth of filamentous algae and fungi (e.g. {\it Phycomyces blakesleeanus}) is associated with a sinistral to dextral reversal~(Castle 1942; Green 1954), a transition in the deformation mode in the fibrillar structure of the cell wall in response to turgor pressure~(Ortega \& Gamow 1974). The sinistral growth occurs via reorientation of initially transverse microfibrils along the growth direction, while the dextral growth immediately following the sporangium formation is dominated by inter-fibril slip~(Ortega \& Gamow 1974) (figure~\ref{fig:natEg}A). The microscopic response is essentially what we see in the uniaxially strained SWCNTs. Control studies have shown that the reversal can also be achieved if the cell wall is allowed to grow laterally~(Yoshida \etal 1980), a combination of the design principles underlying the SWCNT and B-DNA response. 

A similar mechanism may also be involved in the propulsion of {\it Spiroplasma}, a helical wall-less bacteria that infects several plants and insects~(Whitcomb 1980). The undulations that enable it to swim are due to the formation and propagation of kinks, or local reversals in its chirality, along its length~(Shaevitz \etal 2005). The tapered or bulbous ends of the bacteria can facilitate the nucleation the kinks due to motors that generate and then transmit localised axial strains. Another example is the cell wall mediated grain spiral decrease and reversal in drying trunks of conifers (figure~\ref{fig:natEg}B), initiating at the trunk base~(Preston 1950; Kubler 1991). Since the thicker base supports the maximum stress, it likely serves as the origin for changes in cell wall microstructure which eventually express as grain spiral reduction and reversal. In each case, geometry and the imposed strain effect transitions in the interplay between structural chirality and the dominant deformation mode. From a teleological perspective, strain engineering such microstructural transitions is a powerful instrument allowing dynamic control of the coupling. 

We are grateful to L. Mahadevan and Philip Nelson for helpful discussions. MU is also grateful to J. Gore for making a preprint of the paper (Gore \etal 2006) available during the gestation of this study. MU gratefully acknowledges the financial sponsorship from the National Science Foundation (Award CMS-0537537, monitored by Dr. Ken Chong).

%\bibliography{references}

\label{lastpage}
\end{document}